# Characterization of a ballistic supermirror neutron guide

H. Abele[1], D. Dubbers[1], H. Häse[3], M. Klein[1], A. Knöpfler[3], M. Kreuz[2], T. Lauer[1], B. Märkisch[1], D. Mund[1], V. Nesvizhevsky[2], A. Petoukhov[2], C. Schmidt[1], M. Schumann[1], T. Soldner[2]

[1]Physikalisches Institut der Universität, Philosophenweg 12, D-69120 Heidelberg, Germany
[2]Institute Max von Laue-Paul Langevin, BP 156, F-38042 Grenoble Cedex 9, France
[3]S-DH Sputter-Dünnschichttechnik Heidelberg, Hans-Bunte-Str. 8-10, D-69123 Heidelberg, Germany

**Abstract**

We describe the beam characteristics of the first ballistic supermirror neutron guide H113 that feeds the neutron user facility for particle physics PF1B of the Institute Laue-Langevin, Grenoble (ILL). At present, the neutron capture flux density of H113 at its 20×6cm$^2$ exit window is $\Phi_C = 1.35 \cdot 10^{10}$cm$^{-2}$s$^{-1}$, and will soon be raised to above $2 \cdot 10^{10}$cm$^{-2}$s$^{-1}$. Beam divergence is no larger than beam divergence from a conventional Ni coated guide. A model is developed that permits rapid calculation of beam profiles and absolute event rates from such a beam. We propose a procedure that permits inter-comparability of the main features of beams emitted from ballistic or conventional neutron guides.

## 1. Introduction

For many years neutron guides [1] have been used to transport thermal or cold neutrons under small glancing angles almost loss-free from the neutron source to distant neutron instruments. Typical neutron guides have lengths of up to 100m and rectangular cross-sections of area ~100cm$^2$. At first neutron guides were fabricated from glass plates coated on their inside with a thin layer of natural Ni or enriched $^{58}$Ni. In such guides neutrons are transported by total reflection on the inner guide walls, with typical neutron reflection losses of 1% per bounce.

Starting about ten years ago [2-4], neutron guides equipped with 'supermirror' coatings were developed. A neutron supermirror [5] consists of typically 100 double layers of Ni/Ti of varying thickness, and typically doubles the maximum permitted angle of neutron reflection. However, neutron reflection losses are stronger for a supermirror than for a conventional nickel mirror, typically 10% per bounce. Therefore, very long supermirror neutron guides must be of the 'ballistic' type in order to bring benefits. A neutron guide is called ballistic when its cross section varies along its length such as to minimize transport losses within the guide.

In the year 2000 the first ballistic supermirror neutron guide, called H113, of 72m length was installed [6] at the vertical cold neutron source of the Institute Laue-Langevin in Grenoble (ILL). To minimize transport losses, the guide is widened along its initial section from width $2d_0$=6cm to width $2d$=9cm, then the guide has constant width $2d$ along its main central section, and finally the guide is narrowed back to its initial width $2d_0$ in the last section. In this way, the number of neutron reflections on the side walls of the guide is diminished by order $(d_0/d)^2$, and reflection losses are diminished accordingly [6]. The guide feeds the beam station for particle physics PF1B, which is run as a user facility. While in Ref. [6] fabrication and installation of the guide was described, in the present paper we shall describe the properties of the neutron beam emerging from this guide. To this end we develop a simple but effective model, which is able to predict the main properties of lossy neutron guides.

The experiments performed on PF1B need good knowledge of the beam characteristics, both for planning of the experiment and for evaluation of the data. A further reason to publish these characteristics is that at other neutron sources similar beams will become available, and for inter-comparability it will be important to establish some standards of characterization, both for future and for existing guides. Furthermore, experience shows that once such a beam opens up, pressure to start using the beam for experiments is high, and only little time is available for careful studies. Therefore a list of the measurements to be done and pitfalls to avoid may be useful.

2.  **Definitions and pitfalls**

The brightness of a neutron source is

$$B = \frac{\partial^2 \Phi}{\partial \lambda \partial \Omega}, \tag{1}$$

with particle flux density $\Phi$(cm$^{-2}$s$^{-1}$) and solid angle of neutron emission $\Omega$(sterad). Here we call 'source' any neutron emitting surface, like the cold source itself, or the exit window of the neutron guide. In general, brightness $B$ depends on neutron wavelength $\lambda$, on the position $x_0$ on the source, on the direction of neutron emission given by the angles $\theta$ and $\theta'$ in the horizontal and vertical planes, and, for pulsed sources, on time $t$. Here we are concerned only with continuous sources (or with time-average fluxes).

Neutron wavelength $\lambda$ and neutron velocity $\upsilon$ are related by the de Broglie relation $\lambda\upsilon = h/m = 395$nm·ms$^{-1}$. In Eq. (1) wavelength (instead of velocity or energy) is chosen for practical reasons, because neutron spectra usually are measured via time of flight $t$ (over a given distance $z$), which is proportional to wavelength $\lambda$,

$$t(\mathrm{ms}) = \frac{\lambda(\mathrm{nm})\, z(\mathrm{m})}{0.395}. \tag{2}$$

Neutron time-of-flight wavelength spectra are measured with a small pinhole on the source. The emerging neutron beam then is pulsed by a mechanical chopper and registered in a neutron detector further downstream. The chopper gates a time-to-digital converter, which registers the neutron arrival times.

The shape of the measured time-of-flight spectrum may be distorted by various effects, which depend on the following experimental parameters: (i) on the *size* of the detector window; (ii) on the *efficiency* of the detector, determined by the thickness (pressure) of the converter material; (iii) on the detector's *angular position* relative to the axis of the beam passing the pinhole; (iv) on the *time constants* of the detection system. Measurements depend on these parameters in the following ways.

(i)  A *small* detector, with a window of similar size as the pinhole and sufficiently far from it, measures the brightness spectrum $B(\lambda)$. A *large* detector sufficiently near to the pinhole measures the flux density spectrum

$$\frac{\partial \Phi}{\partial \lambda} = \int B\, d\Omega. \tag{3}$$

For an isotropic source, wavelength spectra of brightness $B(\lambda,\theta,\theta')$ and of flux density $\partial\Phi/\partial\lambda$ have the same shape for any $\theta$, $\theta'$. For a strongly anisotropic source like the exit port of a neutron guide, however, wavelength spectra of brightness (including the axial case $\theta=\theta'=0$) and of flux density differ significantly in shape, typically by an additional factor $\lambda^2$ for the latter.

(ii) In a sufficiently *thick* ('black') detector count rate is independent of neutron capture cross section. A thick detector therefore measures the true particle number spectrum $\partial\Phi/\partial\lambda$. In a *thin* detector, count rate is proportional to neutron cross section, which grows linearly with neutron wavelength $\lambda$. A thin detector therefore measures the 'capture-spectrum' $\partial\Phi_C/\partial\lambda$ (also called the 'thermal equivalent spectrum'), which is related to the particle spectrum as

$$\frac{\partial \Phi_C}{\partial \lambda} = \frac{\lambda}{\lambda_0} \frac{\partial \Phi}{\partial \lambda}. \tag{4}$$

$\lambda_0$=0.18nm is the wavelength at the most probable33 velocity $\upsilon_0$=2200ms$^{-1}$ of a Maxwellian 'thermal' spectrum at 300K neutron temperature. For a thermal source with a Maxwellian spectrum, the (wavelength integrated) capture flux density $\Phi_C$, by definition, equals particle flux density $\Phi$. In most experiments on H113 (in the field of nuclear and particle physics) capture flux is the relevant quantity, while in neutron scattering applications it is particle flux. For ILL's cold sources $\Phi_C$ is typically 4 times higher than $\Phi$. All measurements discussed in this article were taken with a 'thin' detector and therefore are of the 'capture' type. They require the suffix C in the formulae of the present section.

A similar distinction should be made for neutron polarization measurements, and in principle one should distinguish also between polarization and 'capture polarization'. On H113, fortunately, this needs not to worry us, because polarization $P_n$=0.997±0.001 and spin flip efficiency >0.999 are both very close to unity for most of the wavelength spectrum [7]. Very long wavelengths ($\lambda$>1nm) where this may not be true can be suppressed by a special supermirror filter [8].

(iii) When a small detector is *off-axis* then the lower part of the wavelength spectrum will be missing if the maximum angle of neutron emission grows with wavelength, as is the case for a neutron guide, see Fig. 2b below.

(iv) Conventional neutron detection systems are *slow* and hence their efficiency must be low enough that they do not saturate at the peak of the spectrum, in particular when flux density spectra, Eq. (3), are measured in a large detector. On the other hand, the use of neutron beam attenuators should be avoided because they distort the wavelength spectrum.

Therefore, measured time-of-flight spectra are not well defined if one or several of the following devices are used:
- a 'gray' detector of medium thickness;
- a detector window of medium size;
- a long narrow slit on the source (chopper) or on the detector;
- a small detector 'off-axis';
- a detector too slow.

In the extreme case of a large thick detector vs. a small thin detector, wavelength spectra from a neutron guide differ by as much as a factor of $\lambda^3$. Therefore published time-of-flight spectra which do not indicate the geometry and type of detector used are of limited value. − We were able to avoid all these pitfalls by using a newly developed thin and fast position sensitive neutron detector to be described in Section 4.

We conclude this list of possible pitfalls with some minor effects. Insufficient absorption by the neutron chopper seems to be a rather trivial source of error. This may happen, though, if, instead of the shortest neutron wavelength occurring in the spectrum, the mean neutron wavelength is used in the exponent of the neutron absorption law. Another source of error is loss of neutrons due to air scattering and absorption. For thermal neutrons, this loss amounts to about 10% per meter flight path and varies with air humidity. To compare, in a helium atmosphere this loss is 0.7% per meter. Finally, for very cold beams and narrow chopper slits, the inner face of the moving chopper slit may hit the neutron and change its direction of flight in a non negligible manner.

Brightness spectra and flux density spectra, Eqs. (1) and (3), will be discussed in Section 5. But the characterization of the neutron beam requires a number of further definitions. The angular distribution of neutrons within the beam is measured with a small detector, or a pixel detector, and a pinhole on the source, as

$$\frac{\partial \Phi}{\partial \Omega} = \int B \, d\lambda. \tag{5}$$

Here brightness $B$ in general depends on the emission angles $\theta$ and $\theta'$. This will be the topic of Sections 6 and 7.

The flux density distribution $\Phi(x_0)$ over the surface of the source is measured with a large detector and a movable pinhole at $x_0$ (alternatively by metal foil activation and image plate read-out [9]) and is, from Eq. (5),

$$\Phi(\mathbf{x}_0) = \iint B \, d\Omega \, d\lambda. \tag{6}$$

Here $d\Omega$ is the solid angle into which the neutrons are emitted from the source. Both $B$ and $d\Omega$ may depend on $\mathbf{x}_0$ and $\lambda$. This will be discussed in Section 8.

The flux density distribution $\Phi(\mathbf{x})$ of a beam downstream of the open source (i.e. without a pinhole on the source) is measured with a small detector at $\mathbf{x}$. (A pixel detector at $\mathbf{x}$ usually would be saturated when used with a large open source.) To calculate $\Phi(\mathbf{x})$ we use the theorem of optical reversibility and write

$$\Phi(\mathbf{x}) = \iint B \, d\Omega^* \, d\lambda. \tag{7}$$

Here integration is over all wavelengths $\lambda$ and over the solid angle $\Omega^*$ under which the source is seen from down-stream position $\mathbf{x}$, where the region of integration $\Omega^*$ in general depends on both $\mathbf{x}$ and $\lambda$. Such neutron beam profiles $\Phi(\mathbf{x})$ will be discussed in Section 8 for an open neutron guide, and in Section 9 for a guide equipped with a collimator system.

Finally, for the planning of experiments total neutron intensity through the instrument is needed. We use the expression 'intensity' instead of the correct expression 'flux', because in neutron physics flux is often used as a synonym for flux density. Neutron intensity is derived by integration of Eq. (7) over a sufficiently large but otherwise arbitrary beam area, which, for a beam sufficiently well collimated along $z$, gives

$$I = \iint \Phi \, dx \, dy, \tag{8}$$

and similarly for the intensity spectrum $\partial I/\partial \lambda$. The total transmission of collimator systems and event rates in real experiments will be discussed in Sections 9 and 10.

The relative importance of the various quantities defined above will depend on the type of instrument installed on the neutron beam. A long instrument usually requires a low-divergence beam along the axis $\theta = \theta' = 0$. If the instrument uses a narrow-band wavelength spectrum centred about a wavelength $\lambda_{ce}$ then the axial brightness $B$ at $\lambda_{ce}$ will be the quantity of interest. If, instead, the instrument uses the full wavelength spectrum, then it will be the on-axis value of $\partial \Phi/\partial \Omega$. For a short instrument, which accepts a high-divergence beam, the quantity of interest will be the spectral flux density $\partial \Phi/\partial \lambda$ at $\lambda_{ce}$ for narrow-band use, and the flux density $\Phi$ for wide-band use.

### 3. Properties of guide H113

We recall some of the properties of the ballistic guide H113, starting with its geometric properties. The guide begins at 2.3m distance from ILL's liquid-deuterium filled vertical cold neutron source of diameter 38cm. It is evacuated to about $10^{-2}$mbar. The guide has a rectangular cross section of constant height $2h_0=20$cm. It consists of several sections:
- A straight 3.5m long section, at present coated with $^{58}$Ni ('$m$=1.2', see below), of nearly constant width (6.8cm at the beginning, 6.0cm at the end). This initial section is followed by 72m of ballistic $m$=2 supermirror guide, consisting of:
- A first 10m long straight section whose width diverges linearly from $2d_0$=6cm to $2d$=9cm.
- An $L$=52m long curved section of constant width $2d$=9cm, with curvature in the horizontal plane and radius $\rho$=4000m; $L$ roughly equals the 'characteristic length' $L_0=8d/\gamma^*=54$m of the curved part of the guide (which is the length below which direct sight through the curved guide becomes possible).
- A final $z'$=10m long straight section whose width converges linearly from $2d$=9cm to $2d_0$=6cm, as shown in Fig. 1.

For the discussion of the surface properties of H113 we recall that in a neutron guide, for a given wavelength $\lambda$, only neutrons with reflection angles $\gamma$ smaller than some limiting angle of reflection $\gamma_c = k_x/k$ are transported. Here $k_x$ is the normal component of the neutron wave vector, and $k=2\pi/\lambda$, the wave number, its length. Hence, this limiting angle is proportional to neutron wavelength, $\gamma_c = \kappa\lambda$, and, for a given angle of reflection $\gamma$, only neutrons with wavelengths $\lambda > \lambda_c = \gamma/\kappa$ are transported. For a single reflecting layer, the reflection mechanism is neutron-optical total reflection, and the 'mirror constant' $\kappa$ is

$$\kappa = k_x / 2\pi = \sqrt{nb/\pi}, \qquad (9)$$

with particle density $n$ and coherent scattering length $b$. Typically, for Ni mirrors, measured mirror constants $\kappa$ are about 10-20% lower than those calculated from Eq. (9), due to a reduced density $n$ of the sputtered or evaporated coating. For a supermirror, the reflection mechanism is Bragg reflection under glancing angles, and $\kappa=1/2a$, where $a$ is the smallest lattice constant in the series of double layers in the coating.

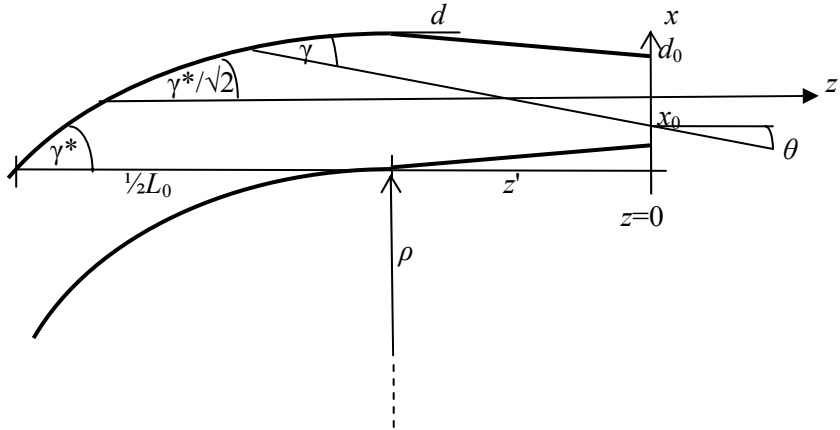

Fig. 1: Horizontal cut of the back end of the ballistic supermirror guide H113 (not to scale). The exit window of the guide is centred at $x_0=0$. A neutron ray is shown which passes the window at $x_0<0$ under an angle of emission $\theta<0$.

The inner wall of the guide H113 is coated with a '$m=2$'-supermirror, which means that the measured mirror constant $\kappa=0.035$rad nm$^{-1}$ is twice the calculated mirror constant of a conventional neutron guide coated with natural nickel, which is 0.0173rad/nm ($m=1$-mirror). For the most likely wavelength of 0.5nm, as read from Fig. 4, the limiting angle then is $\gamma_c=\kappa\lambda=0.0086$. The 'characteristic angle' of the curved part of the guide, cf. Fig. 1, is $\gamma^*=(4d/\rho)^{1/2}=0.0067$.

Neutron reflection losses increase approximately linearly in $\gamma/\gamma_c$, up to the cut-off angle $\gamma=\gamma_c=\kappa\lambda$, that is reflectivity is $R(\gamma)\approx 1-\alpha\gamma/\gamma_c$ for $\gamma\leq\gamma_c$, with loss coefficient $\alpha$, and $R=0$ otherwise, see for instance Fig. 1 in Ref. [6]. At cut-off $\gamma=\gamma_c$, the supermirror coating of H113 has a reflectivity $R(\gamma_c)=88\%$ per bounce. Today mass-produced $m=2$ supermirrors have measured average reflectivity at cut-off of $R(\gamma_c)=91\%$, cf. [10].

At present, the absolute neutron capture flux density on H113 is $\Phi_C=1.35\cdot 10^{10}$cm$^{-2}$s$^{-1}$, as measured by gold foil activation on the exit window of the guide. The 15% drop in flux density, as compared to the value given in [6], is due to a 5m long degraded piece of guide on a BOROFLOAT™ substrate at 15 m distance to the source that will be replaced by a borkron glass substrate in early 2006. At the same occasion, the first 3.5 meters of $^{58}$Ni coated guide will be replaced by a supermirror-coated guide. Both measures are expected to bring flux density to well above $2\cdot 10^{10}$cm$^{-2}$s$^{-1}$. Monte Carlo simulations, performed before the guide was installed, agree with measurements to within several percent, cf. Fig. 2 in Ref. [6], so we think this prediction is reliable.

For an *ideal* $^{nat}$Ni -coating ($m=1$) (roughly the same as a *realistic* $^{58}$Ni coating) of H113, Monte Carlo simulations give a capture flux density of $1.1\cdot 10^{10}$cm$^{-2}$s$^{-1}$. So the gain due to the supermirror coating of H113 will be roughly a factor of two. The gain in capture flux density on H113, as compared to ILL's old particle physics beam station on the conventional $^{58}$Ni guide H53, is even higher (Fig. 2 in Ref. [6]). The additional gain is due to flux-degrading instruments installed upstream on H53, which, by necessity, are absent on a ballistic guide (the source strengths of ILL's two cold sources being equal within about 20% [12]).

## 4. Experimental setup

To measure the full data set $B_C(\mathbf{x}_0,\lambda,\theta,\theta')$ for the guide H113, a pinhole of 1.0×3.1 mm width×height was placed successively at 10 different horizontal and vertical positions on the exit window of the guide, which is at $z=0$ (Fig. 1). Close to this pinhole, 4mm downstream, a neutron absorbing chopper wheel with two open slits each of width 1mm was installed to chop the beam at a wheel radius of 55mm. Spectra were taken with a novel position sensitive neutron detector of 20×20cm$^2$ active area installed at $z=(1698\pm5)$mm distance to the chopper wheel. The beam volume between chopper and detector was kept in a helium atmosphere at standard pressure.

The CASCADE-type neutron detector [11] was used with a single aluminium (100μm thick) drift electrode coated with a $^{10}$B layer of 140nm, thin enough to have a low thermal-neutron sensitivity of 0.74%. Each of the 128×128 pixels, labelled $(\theta,\theta')$, counts neutrons emitted under angles $\theta$ in the horizontal $x$-$z$ plane and $\theta'$ in the vertical $y$-$z$ plane, where $z$ is the axis of the last straight converging section of the guide, cf. Fig. 1. For each pixel a full 14-bit time-of-flight spectrum can be registered. Hence, this detector is an assembly of 16384 'thin' and 'small' detectors in the sense described above, with no blind areas between the pixels, for the direct measurement of brightness spectra $B_C(\lambda,\theta,\theta')$. The dataset of such a measurement contains all relevant information on the neutron beam emerging from a given spot $\mathbf{x}_0=(x_0,y_0,0)$ on the window.

The neutron rate in this dataset was calibrated by normalizing the total number of counts, summed over all channels of the pixel detector, to the measured $\Phi_C=1.35\cdot10^{10}$cm$^{-2}$s$^{-1}$. The calibration of neutron wavelength $\lambda$ and of time zero in Eq. (2) was checked by taking data at two different flight distances $z$.

Beam profiles $\Phi_C(\mathbf{x})$ of the open beam were measured with a small $^3$He detector of low efficiency with a 1mm pinhole on its window.

## 5. Wavelength spectra

We first discuss various neutron *capture brightness spectra* $B_C(\lambda,\theta,\theta')=(\lambda/\lambda_0)B(\lambda,\theta,\theta')$, measured under small solid angle $d\Omega=d\theta\cdot d\theta'$, and then derive the flux density spectrum $\partial\Phi_C/\partial\lambda$ of the guide H113 at the end of this section.

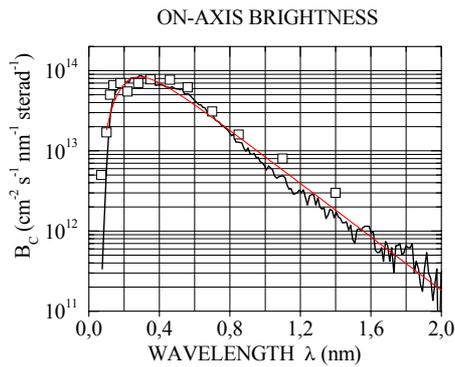 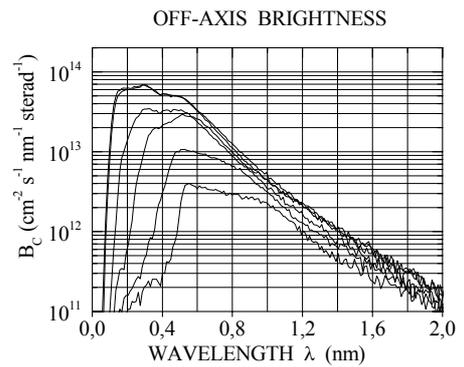

Fig. 2a: Measured absolute on-axis neutron capture-brightness spectrum (black line), at the centre of the exit window of guide H113 ($\theta=\theta'=0$, $\mathbf{x}_0=0$). This spectrum is, within 20%, identical with the original absolute brightness spectrum of the vertical cold source (squares). A fit to Eq. (11) is also given (red line).

Fig. 2b: same as Fig. 2a, but for six different vertical neutron emission angles $\theta'\leq0$ (in multiples of 3.68mrad). The cut-off wavelength $\lambda_c$ increases with increasing $|\theta'|$.

(i) *Cold source*: The absolute capture brightness spectrum $B^{in}_C(\lambda)$ of the vertical cold source is given by the open squares in Fig 2a, with $B^{in}(\lambda)$ taken from [12]. Neutron emission from the cold source is isotropic within the acceptance angle of the guide.

(ii) *Centre of guide exit, along beam axis*: Fig. 2a (line) shows the absolute brightness $B_C(\lambda)$ from a pinhole installed at the center $x_0=0$ of the guide H113, measured on-axis, i.e. under $\theta=\theta'=0$, in the central pixel of the CASCADE detector. This spectrum approximately coincides with the cold source spectrum $B^{in}_C(\lambda)$ 78m upstream measured some 20 years ago. This is not really surprising: Neutrons emitted along the central axis of the beam were last reflected from the guide wall roughly at mid-length of the guide under an angle of reflection $\gamma=\gamma^*/\sqrt{2}$, see Fig. 1. In total these neutrons have made at most two reflections in the guide, and have suffered at most a 20% reflection loss. This means that, when looking through the guide along its axis, one sees, within 20%, the original brightness of the cold source, except for the shortest wavelengths $\lambda<\lambda_c=\gamma/\kappa=0.135$nm for which total reflection is not supported.

(iii) *Centre of guide exit, off-axis*: For non-zero angles of emission $\theta\neq0$ and/or $\theta'\neq0$, the cut-off wavelength $\lambda_c=\gamma/\kappa$ is determined by the largest angle $\gamma$ of total reflection on the guide walls occurring before the emission.

We first discuss a neutron moving in the ($y$, $z$)-plane and leaving the open end of the guide through the central pinhole under angle $\theta'$. Fig. 2b shows the measured brightness spectra for this case, for various angles $\theta'$. For small $|\theta'|$ the largest occurring angle $\gamma$ of total reflection is the constant angle $\gamma^*/\sqrt{2}$ of the central garland reflection on the concave guide wall, with $\lambda_c=\kappa\gamma^*/\sqrt{2}=0.135$nm, independent of $\theta'$. For $|\theta'|>\gamma^*/\sqrt{2}=0.0047$ the largest occurring angle is the reflection angle $\gamma=|\theta'|$ on the flat top and bottom walls of the guide, and $\lambda_c=|\theta'|/\kappa$ becomes linear in $|\theta'|$. This dependence is shown in Fig. 3, where $\lambda_c$ is derived from Fig. 2b at half-maximum of the cut-off slope, with about 5% precision.

The cut-off wavelengths in Fig. 2b behave as expected, with the mirror's $m$ even slightly above the design value $m=2$, see Fig. 3. However, the brightness spectra in Fig. 2b are strongly damped for the larger $\theta'$, because the number of reflections and hence the reflection losses grow with increasing $\theta'$. These losses will be discussed in Section 7.

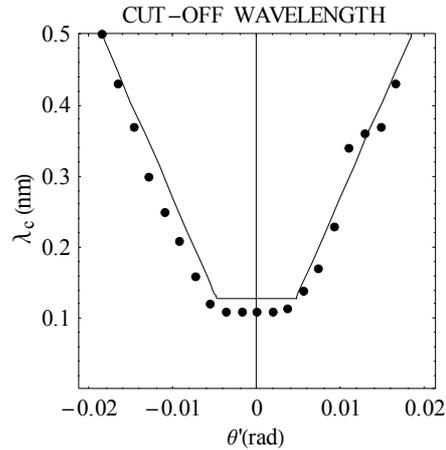

Fig. 3: Neutron cut-off wavelengths $\lambda_c$ (dots) from the measurements of Fig. 2b, and from calculation (line), as a function of the vertical emission angle $\theta'$.

A neutron moving in the ($x$, $z$)-plane and leaving the open end of a curved guide at position $x_0$ under angle $\theta$ was last reflected from the concave guide wall under angle

$$\gamma = \sqrt{\theta^2 + 2(d-x_0)/\rho}. \qquad (10)$$

If the end of the curved guide is seen from a distance $z'$, like in Fig. 1, then $x_0$ in Eq. (10) must be replaced by $x_0+\theta z'$. In a ballistic guide like H113, however, also reflections from the converging pieces at the ends of the guide must be taken into account, and an analytic treatment will not be tried.

(iv) *Guide exit, off-centre*: In a (non-ballistic) curved guide, for emission under $\theta=0$, only garland reflections occur on the concave wall. For garland reflections losses do not depend on the position $x_0$ of the pinhole. This counter-intuitive result is discussed in Section 7. The cut-off wavelength $\lambda_c$, on the other hand, changes with horizontal position $x_0$ by a rather small fraction of approximately $-x_0/2d$, from Eq. (10), where $2d=9$cm is the width of the wide, curved part of the guide. Both features are found to hold within ±10% (not shown).

To simplify further discussion we give an analytical approximation to the measured on-axis brightness $B_C(\lambda, 0, 0)$. Empirically, cold neutron wavelength spectra often are better described by an exponential $\exp(-\lambda/\lambda_1)$, multiplied by some short-wavelength cut-off function, rather than by a Maxwellian spectrum. This is true for the brightness spectra of both cold neutron sources of ILL, for data see Figs. 1, 2 of [12], and also for the brightness spectra of guide H113, cf. Fig. 2. A good parametrization (with no deeper physical meaning) is

$$B_C(\lambda) = B_0 \frac{(\lambda/\lambda_2)^p}{1+(\lambda/\lambda_2)^p} e^{-\lambda/\lambda_1}, \quad (11)$$

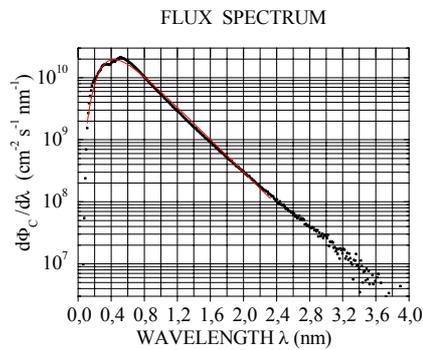

Fig. 4: Absolute neutron capture flux-density spectrum $\partial\Phi_C/\partial\lambda$ averaged over the central $2\times 2$cm$^2$ area of the exit window of the H113 guide, measured (dots) and modelled with Eq. (11) (red line).

For H113, the fit gives, for $\lambda>0.1$nm, the spectral decay constant $\lambda_1=0.26$nm, the cut-off parameters $\lambda_2=0.24$nm and $p=3.0$, and the prefactor $B_0=4\cdot 10^{14}$cm$^{-2}$s$^{-1}$nm$^{-1}$sterad$^{-1}$, see red line in Fig. 2a.

The *capture flux-density spectrum* $\partial\Phi_C/\partial\lambda$, Eqs. (3) and (4), is obtained by adding up all spectra from all channels of the pixel detector, see dots in Fig. 4. For $\lambda>0.1$nm the spectrum again follows the parametrization Eq. (11), with $\lambda_1=0.33$nm, $\lambda_2=0.40$nm, $p=3.0$, and a pre-factor $1.3\cdot 10^{11}$cm$^{-2}$s$^{-1}$nm$^{-1}$, see the red line in Fig. 4. From this spectrum we can calculate the *particle flux density* to $\Phi(0)=5.0\cdot 10^9$cm$^{-2}$s$^{-1}$, which is 2.7 times lower than the capture flux density of $\Phi_C(0)=1.35\cdot 10^{10}$cm$^{-2}$s$^{-1}$. For low-loss Ni coated cold guides like the guide H53, particle flux is lower than capture flux by a factor between 3.5 and 4, which indicates that the spectrum from H113 is warmer than a typical cold spectrum, as expected (see Fig. 3 of Ref. [13]). Therefore, gain in particle flux density on H113 is higher by a factor between 1.3 and 1.5 than gain in capture flux density. As compared to the $^{58}$Ni guide H53, the gain in particle flux density on H113 therefore will be about a factor seven (which, again, only in part is due to the supermirror coating, as stated at the end of Section 3).

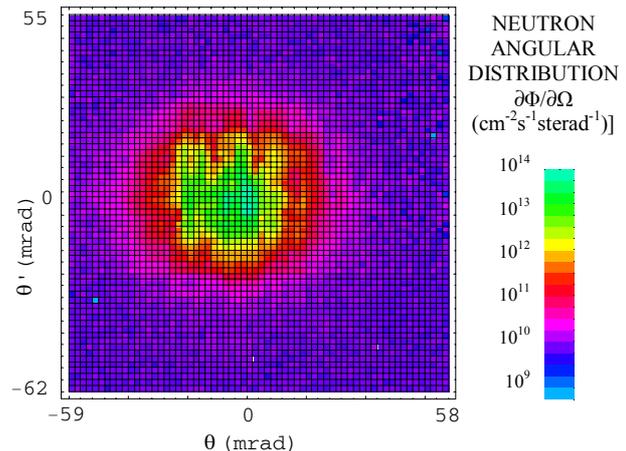

Fig. 5: Neutron angular distribution measured with a CASCADE detector at 1.7m distance from a pinhole at the center $x_0=0$ of the exit window of the ballistic guide H113. Data are displayed in 64×64 channels. Each channel in this measurement holds a 9-bit time-of-flight spectrum. The small asymmetry in the horizontal distribution appears enhanced due to the logarithmic presentation of the data.

## 6. Angular distributions

Fig. 5 shows a contour plot of the neutron angular distribution $\partial \Phi_C/\partial \Omega$, Eq. (5), from the central pinhole, in dependence of emission angles $\theta$ and $\theta'$. The detector simply acts as a 'pinhole camera', which images the interior of the guide. Like the guide itself, the horizontal distribution is more structured than the vertical distribution.

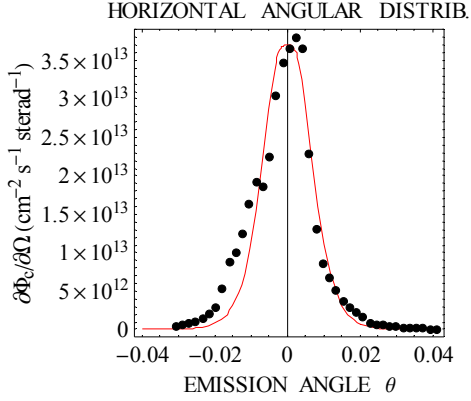 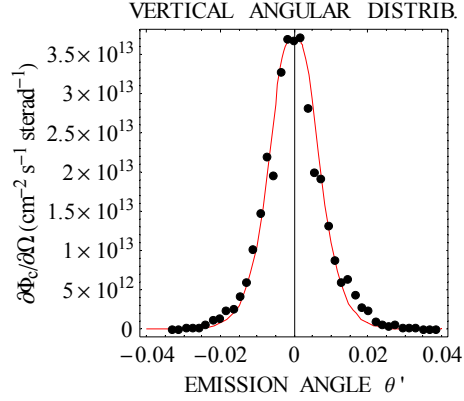

Fig. 6a: Measured horizontal angular distribution of the flux density from a central pinhole, and fit to Eqs. (12) plus (11), with $\kappa_{eff}=0.017$ rad/nm.

Fig. 6b: Same as Fig. 6a, but for vertical angular distribution.

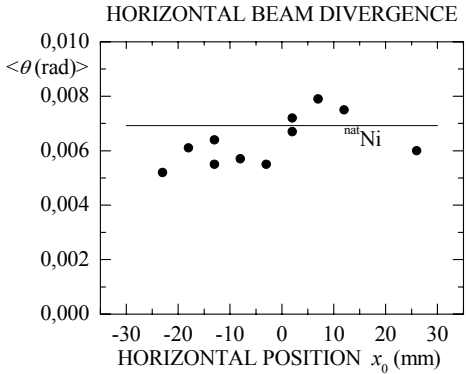 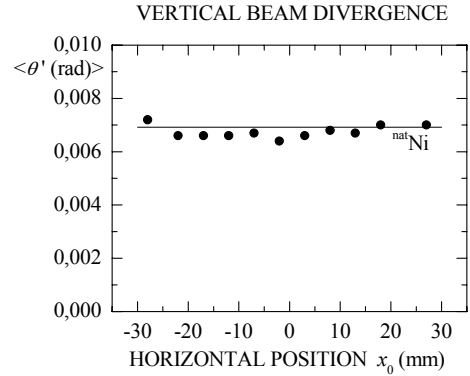

Fig. 7a: Horizontal divergence $<\theta>$ of the H113 neutron beam, as a function of horizontal position $x_0$ on the guide exit. The mean beam divergence of a conventional cold $^{nat}$Ni guide is added for comparison (horizontal line).

Fig. 7b: Same as Fig. 7a, but for the vertical divergence $<\theta'>$. The good agreement with a $^{nat}$Ni guide is accidental.

Figs. 6a and 6b show horizontal and vertical cuts through the centre of Fig. 5. The divergence (half width at half maximum from Figs. 6) is $<\theta>\approx<\theta'>=0.007$. The horizontal divergence $<\theta>$, which is reduced during passage of the first diverging piece of the guide, hence is restored after passage of the last converging section of the guide to the same value as the vertical divergence, which latter, for $\theta=0$, is hardly affected by the 'ballistic' structure of the guide. As shown in Figs. 7a and 7b, the vertical divergence $<\theta'>$ varies only little over the surface of the guide exit, while the horizontal divergence $<\theta>$ is less stable, due to the more complicated structure of the ballistic guide in the horizontal plane, cf. Fig. 1.

By pure accident, the divergence of the H113 beam is found roughly the same as that expected for a straight guide with a $^{nat}$Ni coating. This is consistent with the analytical result [13] that a supermirror coating of a (non-ballistic) neutron guide does not lead to a noticeable increase in beam divergence, contrary to popular belief. Monte Carlo simulations for the ballistic H113 guide support this result, giving $<\theta>\approx 0.0075$ and $<\theta'>\approx 0.0065$, each with ~10% statistical error.

So the intensity gain achieved with a long supermirror guide is only to a smaller part due to its higher angle of acceptance, because the corresponding trajectories are damped out by reflectivity losses. To a larger part, the gain is due to the fact that, at a given angle of reflection $\gamma$, a supermirror transports neutrons with shorter wavelengths than a conventional Ni mirror. Hence a cold beam from a supermirror guide is warmer than a cold beam from a Ni guide, as stated at the end of the preceding section, but this effect is already accounted for in the measured capture flux density $\Phi_C$.

## 7. Modelling of lossy guides

In order to understand qualitatively the properties of a lossy ballistic guide, we shall discuss the transmission function of various lossy guides, including straight and curved conventional guides. Here 'conventional guide' here means a non-ballistic guide, i.e. with constant rectangular cross-section. We then develop a realistic model for a lossy ballistic guide.

(i) For a *lossless guide of any shape* the angular distribution of the flux density from a pinhole is, from Eq. (5),

$$\frac{\partial \Phi(\theta,\theta')}{\partial \Omega} = \int_{\lambda_c(\theta,\theta')}^{\infty} B^{in}(\lambda)\, d\lambda. \tag{12}$$

(If needed, the suffix C can be added.) Here $B^{in}(\lambda)$ is the isotropic brightness of the source feeding the guide given in (i) of Section 5, which is also well modelled by Eq. (11). $B^{in}(\lambda)$ is independent of position on the source and emission angles $\theta$ and $\theta'$. Integration starts at the $(\theta,\theta')$-dependent cut-off wavelength $\lambda_c = \gamma/\kappa$. Here, for a given neutron ray, $\gamma(\theta,\theta')$ is the largest occurring angle of total reflection on the walls of the guide which, in general, also depends on $x_0$.

In the special case of a long conventional *straight guide*, $\lambda_c = \max(|\theta|, |\theta'|)/\kappa$, and evaluation of Eq. (12) for a given $B^{in}(\lambda)$ and $\theta, \theta'$ is straightforward. For a conventional *curved guide*, the reflection angle $\lambda_c = \gamma(\theta)/\kappa$ in the horizontal is given by Eq. (10), while in the vertical $\lambda_c = |\theta'|/\kappa$.

(ii) For a *lossy guide*, $B^{in}(\lambda)$ in Eq. (12) must be multiplied with a transmission function such that $B(\lambda,\theta,\theta') = T(\lambda,\theta,\theta')B^{in}(\lambda)$. For $N$ reflections under angle $\gamma$, the transmission function is best parametrized as $T = R^N = \exp(-N\alpha\gamma/\gamma_c)$, with $\gamma_c = \kappa\lambda$. For a *lossy straight guide* of cross-section $2d\times 2h$, the number of reflections under angle $\gamma = |\theta|$ is $N = |\theta|L/2d$ for the side walls, and $N' = |\theta'|L/2h$ for the top and bottom walls, so $T_{straight}(\lambda,\theta,\theta') = T_{hor}(\lambda,\theta)\, T_{vert}(\lambda,\theta')$, with

$$T_{vert}(\lambda,\theta) = \exp\left(-\alpha \frac{L}{2h}\frac{\theta'^2}{\kappa\lambda}\right), \tag{13}$$

and similarly for $T_{hor}(\lambda,\theta)$, with $h$ replaced by $d$. Here and in the following we assume that the transmission function is separable in $\theta$ and $\theta'$, i.e. has the form $f(\theta)\cdot g(\theta')$, as expected for a guide of (even varying) rectangular cross-section and of uniform quality. From Eq. (13) we see that losses increase with decreasing $\lambda$, and strongly increase with increasing $\theta$ and $\theta'$. This general trend can also be seen in the data from a ballistic guide, see Fig 2b.

(iii) In a conventional *lossy curved guide*, for reflections under angles $\gamma < \gamma^*$ (in the *horizontal* plane $\theta'=0$), only 'garland reflections' occur [1], which involve only the outer concave guide wall. Their number is $N = (L/L_0)\gamma^*/\gamma$, and transmission

$$T_{\text{hor, garland}} = \exp\left(-\alpha \frac{L}{L_0} \frac{\gamma^*}{\kappa\lambda}\right), \tag{14}$$

becomes independent of reflection and emission angles $\theta$, $\gamma$, and of position $x_0$ on the exit window. With $\alpha$=0.12, we find $T_{\text{hor, garland}}$=exp(−0.055/$\lambda$(nm)) for the curved section of H113. For $\gamma \geq \gamma^*$ and $\theta'$=0, only '*zigzag reflections*' occur, which involve also the inner convex guide wall, and transmission becomes

$$T_{\text{hor, zigzag}} = \exp\left(-\alpha \frac{L}{L_0} \frac{\gamma^*}{\kappa\lambda}\left(\gamma/\gamma^* + \sqrt{\gamma^2/\gamma^{*2} - 1}\right)^2\right), \tag{15}$$

which depends, via Eq. (10), on $\theta$ and $x_0$.

For emission under angles $\theta' \neq 0$ in the *vertical* plane also the upper and lower flat walls of the curved guide become involved, and the above transmission factors must be multiplied by $T_{\text{vert}}$ from Eq. (13). Hence, for a lossy curved conventional guide the transmission function for both the brightness $B^{\text{in}}(\lambda)$ and the flux density spectrum $\partial\Phi/\partial\lambda$ [13] can be given in closed form.

(iv) For a *lossy curved ballistic guide* (i.e. of varying cross-section) the transmission function can, in general, not be given in closed form. However, we shall develop a simple procedure that permits to model the relevant properties of the beam from a lossy ballistic guide with good precision. In our model we shall still use Eq. (12) with the undisturbed on-axis brightness $B^{\text{in}}(\lambda)$, like in the case of a lossless guide. Further, we shall use the cut-off wavelength $\lambda_c = |\theta|/\kappa$ (for $\theta \geq \theta'$), like in the case of a conventioanl straight guide. We merely replace the mirror constant $\kappa$ by some effective value $\kappa_{\text{eff}} < \kappa$, i.e. we use $\lambda_{c,\text{eff}} = |\theta|/\kappa_{\text{eff}}$, and similarly for $\theta'$ with an effective constant $\kappa_{\text{eff}}'$.

At first sight the introduction of an effective mirror constant $\kappa_{\text{eff}}$ does not seem to be a good approximation, because Figs. 2b and 3 show that the brightness spectrum *does* start at cut-off $\lambda_c$, and not at some effective value $\lambda_{c,\text{eff}} > \lambda_c$. But our expectation is that, when a properly chosen $\kappa_{\text{eff}}$ is used with $\lambda_c$ in Eq. (12), this may account for the losses in $B_C(\lambda)$ in a roughly area preserving way, when, instead of decreasing $B_C(\lambda)$, we increase $\lambda_c$ in Fig. 2b appropriately. The curves in Figs. 6a and 6b are fits to Eq. (12), using $B_C$ from Eq. (11) for $B^{\text{in}}$, with the values $\lambda_1$, $\lambda_2$, $p$ quoted there, and $\kappa_{\text{eff}}$ and $B_0$ as fit parameter. The curves in Fig. 6 can be understood as being composed of infinitesimal areas of width $2\kappa_{\text{eff}}\lambda$ and height $B_C(\lambda)d\lambda$. From Fig. 6a we find the same value $\kappa_{\text{eff}}$=(0.017±0.002) rad/nm as from Fig. 6b, $\kappa_{\text{eff}}'$=(0.017±0.001) rad/nm, though with a somewhat larger error. This, as mentioned above, is close to the theoretical mirror constant of a $^{\text{nat}}$Ni coating.

The fits in Figs. 6 indicate that the concept of an effective mirror constant $\kappa_{\text{eff}}$ is reasonable: If agreement is acceptable for the elementary angular distribution from a pinhole, then it will also be acceptable for global beam profiles, and also for event rates derived by integration over detector and source areas. Indeed, after multiple integration some of the remaining irregularities may even be washed out.

To give an example for this statement: Our model assumes that the neutrons from a pinhole are emitted isotropically into a rectangular cone of solid angle $\Delta\Omega = 4\kappa_{\text{eff}}^2\lambda^2$. Therefore the integrated capture flux density should be, from Eq. (6),

$$\Phi_C(x_0 = 0) = 4\kappa_{\text{eff}}^2 \int_0^\infty \lambda^2 B_C(\lambda) d\lambda, \tag{16}$$

and should agree with the measured capture flux density $\Phi_C(0)$. In fact, with $B_C(\lambda)$ from Eq. (11) used in Eq. (16), we find $\Phi_C(0)$=1.43·10$^{10}$cm$^{-2}$s$^{-1}$. This rather good agreement with the measured value $\Phi_C(0)$=1.35·10$^{10}$cm$^{-2}$s$^{-1}$ strengthens our confidence in the model.

Please note: while, in our model, the integral Eq. (16) over $\lambda$ does give the correct flux density $\Phi_C(0)$, the integrand in Eq. (16) itself needs not be equal to the flux density spectrum $\partial\Phi_C/\partial\lambda$, which, indeed, we know to be warmer than the flux density spectrum $4\kappa^2\lambda^2 B(\lambda)$ from a lossless straight guide.

## 8. Beam profiles

Next we are interested in the flux density profiles $\Phi_C(x)$ of the neutron beam delivered from the *open exit window* of the guide (i.e. without pinhole). From Eq. (7), these flux density profiles have the shape, in dependence of position $x=(x, y, z>0)$,

$$\Phi_C(x) = \int_0^\infty B_C(\lambda) \delta_0(\lambda,x,z) \delta_0'(\lambda,y,z) \mathrm{d}\lambda, \qquad (17)$$

where $\delta_0$ is the opening angle in the $(x, z)$ plane under which the neutrons emerging from the guide exit are seen from position $x$. The opening angle $\delta_0$ is limited either by $\kappa_{eff} \lambda$, or by the edges of the guide's window, i.e.

$$\delta_0(\lambda,x,z) = \min(\kappa_{eff}\lambda, (d_0-x)/z) - \max(-\kappa_{eff}\lambda, -(d_0+x)/z), \qquad (18)$$

and correspondingly for the vertical opening angle $\delta_0'(\lambda,y,z)$. In Eq. (18) we used our model assumption that, for a given wavelength $\lambda$, the flux is isotropic within $|\theta| \leq \kappa_{eff} \lambda$ and $|\theta'| \leq \kappa_{eff}' \lambda$, and is uniform over the surface $2d_0 \times 2h_0$ of the guide exit, cf. Fig. 8b.

As an example, Fig. 8a gives the calculated capture flux density $\Phi_C(z)/\Phi_C(0)$ on-axis ($x=y=0$) as a function of distance $z$ from the exit of the open guide, using $B_C$ from Eq. (11) with the parameters given there. Fig. 8b shows the measured and calculated horizontal flux density profile $\Phi_C(x)/\Phi_C(0)$ of H113 at a distance $z=0.5$m from the guide window. At 0.5m distance the beam diverges only by $2z\kappa_{eff}<\lambda>\sim 7$mm, so we can conclude from Fig. 8b that the flux density over the exit window of H113 is uniform to about 5%, in agreement with activation measurements on different locations on the guide window. The slight increase of the measured flux density towards the outer concave face of the guide is what we expect from the garland reflections. − The vertical flux variations are even smaller.

For a uniform beam profile the total capture intensity of H113 then is, from Eq. (8),

$$I_C = 4 d_0 h_0 \Phi_C(0) = 2.0 \cdot 10^{12} \, \mathrm{s}^{-1}. \qquad (19)$$

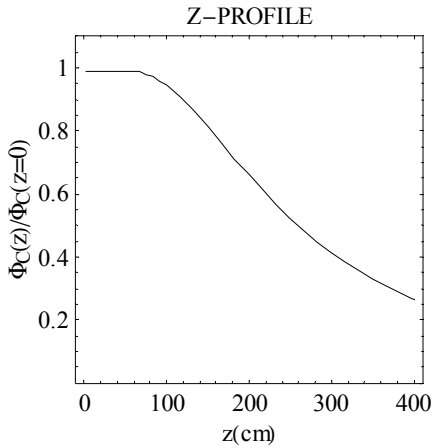
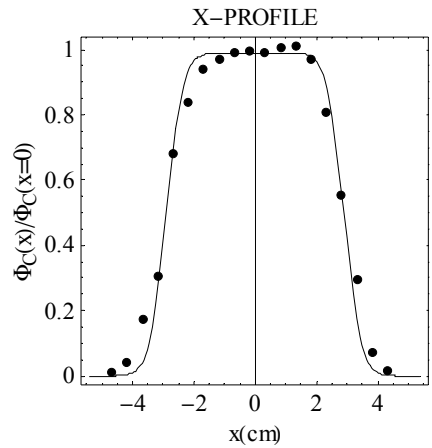

Fig. 8a: Neutron flux density $\Phi_C(z)/\Phi_C(0)$ as a function of distance $z$ from the exit of the open guide, on-axis $x=y=0$, from Eqs. (17) and (18).

Fig. 8b: Measured horizontal neutron beam profile $\Phi_C(x)/\Phi_C(0)$ (dots), and expectations (line) from Eq. (17), at mid beam-height $y=0$ and distance $z=0.5$m.

## 9. Transmission of a collimator system

We want to test the predictive power of our model by calculating the total event rate in a recent experiment on free neutron decay. To this end we must slightly extend our treatment. In all experiments done on H113 (or elsewhere) the neutrons first have to pass a collimator system. The collimator usually consists of a series of *n* rectangular collinear orifices of widths $2d_i$ and heights $2h_i$, placed at distances $z_i$ to the source, see Fig. 9 for a single orifice $n=1$.

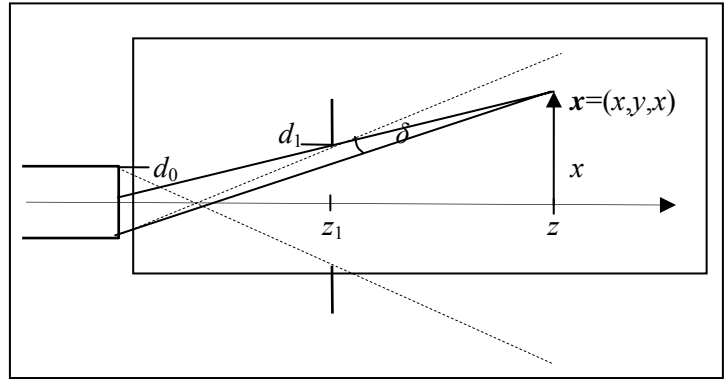

Fig. 9: View through a single orifice onto the port of a neutron guide.

To obtain the new opening angles $\delta_n$ under which the source is seen through the *n* orifices we merely have to add to the opening angles in Eq. (18) *n* additional angles to obtain

$$\delta_n(\lambda, x, z) = \min(\kappa_{\text{eff}}\lambda, \{(d_i - x)/(z - z_i); i = 0,1,...,n\}) \\ - \max(-\kappa_{\text{eff}}\lambda, \{-(d_i + x)/(z - z_i); i = 0,1,...,n\}), \quad (20)$$

with $z_0=0$, and the same for $\delta_n'(\lambda, y, z)$. Negative $\delta_n$ or $\delta_n'$ are set to zero. These angles are very simple to evaluate numerically and allow derivation of all relevant spectra and profiles in a straightforward way. None of the quantities derived in this article require more that three lines of code in a high level mathematical language. In this way one can avoid that every user of the beam station has to run a special Monte Carlo program to characterize the setup in use. The collimator system used in the experiment consisted of six orifices [9] of

$$\begin{aligned}
\text{widths} \quad & 2d_i(\text{cm}) = \{5.80,\ 5.10,\ 4.25,\ 3.33,\ 3.17,\ 3.23\}; \\
\text{heights} \quad & 2h_i(\text{cm}) = \{3.75,\ 4.20,\ 3.05,\ 2.23,\ 1.83,\ 1.89\}; \\
\text{positions} \quad & z_i(\text{m}) = \{0.00,\ 0.22,\ 0.39,\ 1.34,\ 2.14,\ 2.67\}.
\end{aligned} \quad (21)$$

The flux density profiles $\Phi_{\text{coll}}(x)$ behind the collimator then are again given by Eq. (17), with $\delta_0$, $\delta_0'$ replaced by $\delta_n$, $\delta_n'$ from Eq. (20). As an example Fig. 10 gives the calculated lateral beam profile at distance $z=3$m with and without collimator. Division of both profiles gives the transmission of the collimator $T(x)=\Phi_{\text{coll}}(x)/\Phi(x)$, spatially resolved (not shown).

For all experiments it is important to know the total neutron intensity in order to predict experimental event rates. The total neutron intensity $I_{\text{coll}}$ from the guide plus collimator is (for the moment we drop the subscript C)

$$I_{\text{coll}} = \int_0^\infty \frac{\partial I_{\text{coll}}}{\partial \lambda} d\lambda, \quad (22)$$

with the intensity spectrum, from Eqs. (8) and (3),

$$\frac{\partial I_{\text{coll}}}{\partial \lambda} = B(\lambda) \int d\Omega\, dx dy = B(\lambda) F(\lambda) F'(\lambda), \quad (23)$$

where brightness $B(\lambda)$ is isotropic within $\pm\kappa_{\text{eff}}\lambda$, and where the $F(\lambda)$ are integrals over the area $2d_n \times 2h_n$ of the last orifice at position $z_n$ (or, as total intensity is conserved, over any other large area beyond $z_n$)

$$F(\lambda) = \int_{-d_n}^{+d_n} \delta_{n-1}(\lambda, x, z_n)\, dx, \quad F'(\lambda) = \int_{-h_n}^{+h_n} \delta_{n-1}'(\lambda, y, z_n)\, dy, \quad (24)$$

with $\delta_{n-1}$ and $\delta_{n-1}'$ as defined in Eq. (20), but for $n-1$ orifices (here as seen from the last orifice at $z=z_n$).

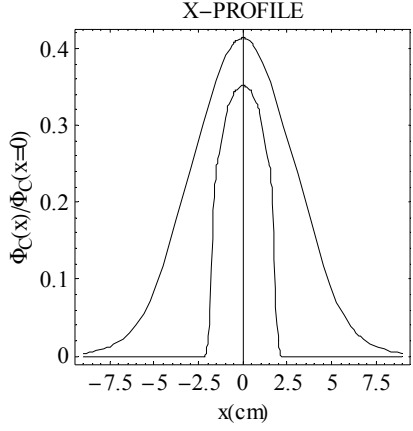
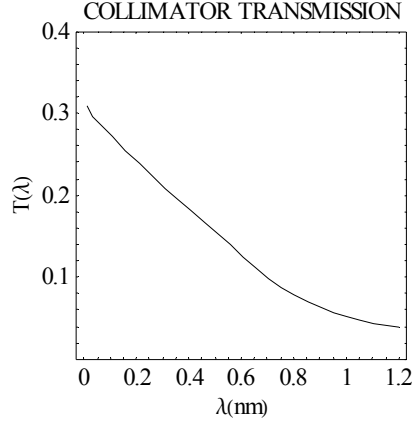

Fig. 10: Horizontal flux density profiles $\Phi_C(x)/\Phi_C(0)$, Eq. (17), with and without collimator (Eqs. (18) and (20)), at mid beam-height $y=0$ and distance $z=3$m.

Fig. 11: Wavelength dependent total transmission $T(\lambda)$ of collimator system, Eqs. (26) and (23), at distance $z=3$m.

For a single orifice $n=1$ these integrals can be given in closed form:

$$F(\lambda) = 4\kappa\lambda d_1 \qquad \text{for } 0 \le \kappa\lambda < (d_0 - d_1)/z_1,$$
$$F(\lambda) = 4\kappa\lambda d_1 - (\kappa\lambda z_1 - d_0 + d_1)^2/z_1 \qquad \text{for } (d_0 - d_1)/z_1 \le \kappa\lambda < (d_0 + d_1)/z_1, \qquad (25)$$
$$F(\lambda) = 2d_0 d_1/z_1 \qquad \text{for } (d_0 + d_1)/z_1 \le \kappa\lambda < \infty,$$

and similarly for $F'(\lambda)$. For $n>1$, the integrals Eq. (24) are easily summed up numerically.

With the same arguments as for the flux density, we note that with our approximation the integral Eq. (22) for the total intensity is reliable, but the integrand Eq. (23) for the intensity spectrum is much less so. On the other hand, the transmission spectrum of the collimator

$$T(\lambda) = \frac{\partial I_{\text{coll}}/\partial \lambda}{\partial I/\partial \lambda}, \qquad (26)$$

with neutron intensity $I$ of the guide without collimator, will be somewhat more reliable because spectral distortions will tend to shorten out. Fig. 11 shows the calculated spectral transmission $T(\lambda)$ of the collimator system at distance $z=3$m. Transmission is higher for neutrons with short wavelengths, which have a smaller divergence $<\theta>=\kappa_{\text{eff}}\lambda$ and hence are less affected by the collimator. The global transmission $T$ of the collimator

$$T = I_{\text{coll}}/I, \qquad (27)$$

on the other hand, should again be well reliable. − These formulae can be used to rapidly optimize the geometry of an experimental installation for maximum neutron intensity.

## 10. Event rates

The neutron intensities derived in the foregoing section can be used to predict event rates in the planning of experiments. As an example, we shall use the formalism to predict the neutron decay rate from a recent experiment on H113. The in-beam neutron decay rate from an active volume of length $\ell$ of a neutron beam of capture intensity $I_C$ is

$$n_\beta = \frac{\ell}{v_0 \tau} I_C, \qquad (28)$$

with neutron lifetime $\tau=886$s and $v_0=2200\text{ms}^{-1}$.

In the experiment the decay of *polarized* neutron was measured, with two crossed $m=2.8$ supermirror polarizers [7] installed behind the guide exit, each of length 80cm, radius of curvature 30m, and channel width 2mm. The flux density at the exit of the polarizers is $\Phi_C(0)=1.6\cdot10^9\mathrm{cm}^{-2}\mathrm{s}^{-1}$, measured by gold foil activation. The divergence $<\theta_c>=0.007$ (Fig. 7) of the neutrons entering the polarizer is smaller than the characteristic angle $\gamma^*_{\mathrm{pol}}=0.012$ of the polarizer channels. Hence most neutrons will be transported through the polarizer by garland reflections, in which case the divergence of the neutrons leaving the polarizer will be (from Eq. (10), for neutrons entering a polarizer channel near its centre $x_0=0$)

$$<\theta_{c,\mathrm{pol}}> \approx \sqrt{<\theta_c>^2 + \gamma^{*2}_{\mathrm{pol}}/2} = 0.0108 = 1.53\cdot<\theta_c> \qquad (29)$$

Hence we shall use $\kappa_{\mathrm{eff,pol}}=1.53\cdot\kappa_{\mathrm{eff}}=0.026\mathrm{nm}^{-1}$. The polarizer's $m=2.8$ had been chosen 1.4 times larger than the H113 guide's $m=2$ in order to cope with these larger reflection angles.

It turns out that the measured polarized spectrum on-axis can again be modelled with Eq. (11) over the whole wavelength spectrum, using $\lambda_1=0.19\mathrm{nm}$, $\lambda_2=0.46\mathrm{nm}$, $p=3.4$, and, using Eq. (16), $B_0=8.2\cdot10^{13}\mathrm{cm}^{-2}\mathrm{s}^{-1}\mathrm{nm}^{-1}\mathrm{sterad}^{-1}$. With $\kappa_{\mathrm{eff}}=0.026\mathrm{nm}^{-1}$, the transmission of the collimator system Eq. (27) for this spectrum is $T_C=6.5\%$. For a length of the decay volume $\ell=0.27\mathrm{m}$ we then obtain the neutron decay rate is $n_\beta=319\mathrm{s}^{-1}$. The measured neutron decay rate is $n_\beta=388\mathrm{s}^{-1}$ [9]. In view of the crudeness of our treatment of the polarizer, this agreement is satisfactory.

## 11. Summary

The on-axis brightness of the ballistic supermirror cold-neutron guide H113 is, within about 20%, the same as the brightness of the cold source 78m upstream. The mean divergence of the beam of ±7mrad is similar to the divergence from a conventional Ni coated cold guide. Hence, even low-divergence experiments can fully profit from the gain in neutron flux density provided by the ballistic supermirror guide H113. When a fast position sensitive neutron detector like the newly developed CASCADE detector is used for beam characterization, rapid data collection is possible, while the usual pitfalls in neutron beam characterization can be avoided.

We believe that the simple model developed in this article is also applicable to other neutron guides. Therefore, to make the characteristics of different neutron guides inter-comparable, we propose to indicate, besides the capture flux density $\Phi_0$ and the area $2d_0\times2h_0$ of the exit window of the guide, the following quantities.
  (i) In the parametrization of Eq. (11), the spectral decay constants $\lambda_1$, the cut-off parameters $\lambda_2$ and $p$, and the absolute pre-factors both for the on-axis brightness $B(\lambda)$ and the flux density spectrum $\partial\Phi/\partial\lambda$.
  (ii) In the model described under (iv) in Section 7, the values for effective mirror constants $\kappa_{\mathrm{eff}}$ and $\kappa_{\mathrm{eff}}'$.
  (iii) The flux variation over the entrance window.

With the given parameters $\kappa_{\mathrm{eff}}$, $\kappa_{\mathrm{eff}}'$, $\lambda_1$, $\lambda_2$, $p$, $B_0$, all beam profiles Eq. (17) with (18) and (20), spectral transmission functions Eq. (26), and event rates like Eq. (28), can then be predicted for any collimator system of the type given in Eq. (21). A set of such numbers therefore uniquely characterizes a given neutron guide system, ballistic or conventional. For ILL's guide H113, these numbers are $\kappa_{\mathrm{eff}}=\kappa_{\mathrm{eff}}'=0.017\mathrm{rad/nm}$, and $\lambda_1=0.26\mathrm{nm}$, $\lambda_2=0.24\mathrm{nm}$, $p=3.0$, $B_0=4\cdot10^{14}\mathrm{cm}^{-2}\mathrm{s}^{-1}\mathrm{nm}^{-1}\mathrm{sterad}^{-1}$ for the on-axis brightness $B_C(\lambda)$. The flux density spectrum $\partial\Phi_C/\partial\lambda$ has, for the same parametrization, the values $\lambda_1=0.33\mathrm{nm}$ and $\lambda_2=0.40\mathrm{nm}$.

This work was funded by the German Federal Ministry for Research and Education under Contract No. 06HD153I.

**References**

[1] H. Maier-Leibnitz, T. Springer, Reactor Sci. and Technol. **17**, 217 (1963)
[2] U. Schmidt, D. Dubbers, K. Raum, O. Joeres, O. Schaerpf,
 J. Neutron Research **5**, 81 (1996)
[3] K. Al Usta, R. Gäehler, P. Böni, P. Hank, R. Kahn, M. Koppe, A. Menelle, W. Petry,
 Physica **B241**, 77 (1998)
[4] S. Itoh, T. Kamiyama, M. Furusaka, S. Ikeda, Physica **B241**, 79 (1998)
[5] F. Mezei, Comm. on Phys. **1**, 81 (1976)
[6] H. Häse, A. Knöpfler, K. Fiederer, U. Schmidt, D. Dubbers, W. Kaiser,
 Nucl. Instr. Meth. **A485**, 453 (2002)
[7] M. Kreuz, V. Nesvizhevsky, A. Petoukhov, T. Soldner,
 Nucl. Instr. Meth. **A547**, 583 (2005)
[8] P. Høghøj, H. Abele, M. Astruc Hoffmann, S. Baeßler, J. Reich, V. Nesvizhevsky,
 O. Zimmer, Nucl. Instr. Meth. **B160**, 431 (2000)
[9] M. Schumann, diploma thesis, Heidelberg 2004, unpublished
[10] http://www.s-dh.de/
[11] M. Klein et al., to be published
[12] P. Ageron, Nucl. Instr. Meth. **A284**, 197 (1989)
[13] D. Dubbers, Nucl. Instr. Meth **A349**, 302 (1994)